\documentclass[twocolumn]{revtex4}
\usepackage{multirow}
\usepackage{pstricks}
\usepackage{dcolumn}
\usepackage{bm}
\usepackage{latexsym}
\usepackage{dcolumn}
\usepackage{amsmath}
\usepackage{amsfonts,amssymb}
\usepackage{graphicx,epsfig}
\usepackage{color}
\usepackage{psfrag}
\usepackage{amsthm}
\usepackage{ulem} 
\usepackage{soul} 
\usepackage{flushend}
\usepackage{titlesec}

\def\be{\begin{equation}}
\def\ee{\end{equation}}
\def\be{\begin{equation}}
\def\ee{\end{equation}}
\def\bea{\begin{eqnarray}}
\def\eea{\end{eqnarray}}

\def\bes{\begin{equation*}}
\def\ees{\end{equation*}}
\usepackage{slashed}

\def\lsim{\raise0.3ex\hbox{$\;<$\kern-0.75em\raise-1.1ex\hbox{$\sim\;$}}}
\def\gsim{\raise0.3ex\hbox{$\;>$\kern-0.75em\raise-1.1ex\hbox{$\sim\;$}}}

\begin{document}
\begin{flushleft}
SHIP-HEP-2020-01 \\
\end{flushleft}
\title{Probing $Z^\prime$ Mediated Charged Lepton Flavor\\ Violation with Taus at the LHeC}
\author{Stefan Antusch$^{\dagger}$, A. Hammad$^{\dagger}$ and Ahmed Rashed$^{\ddagger}$}
\affiliation{
$^\dagger$Department of Physics, University of Basel, Klingelbergstr.\ 82, CH-4056 Basel, Switzerland\\
$^{\ddagger}$ Department  of Physics,  Shippensburg University of Pennsylvania,\\
 Franklin Science Center, 1871 Old Main Drive, Pennsylvania, 17257, USA}

\begin{abstract}
\noindent While charged lepton flavor violation (cLFV) with taus is often expected to be largest in many extensions of the Standard Model (SM), it is currently much less constrained than cLFV with electrons and muons. We study the sensitivity of the LHeC to $e$-$\tau$ (and $e$-$\mu$) conversion processes $p  e^- \to \tau^- + j$ (and $p  e^- \to \mu^- + j$) mediated by a $Z'$ with flavor-violating couplings to charged leptons in the $t$-channel. Compared to current tests at the LHC, where cLFV decays of the $Z'$ (produced in the s-channel) are searched for, the LHeC has sensitivity to much higher $Z'$ masses, up to {\cal O}(10) TeV. For cLFV with taus, we find that the LHeC sensitivity from the process $p  e^- \to \tau^- + j$ can exceed the current limits from collider and non-collider experiments in the whole considered $Z'$ mass range (above $500$ GeV) by more than two orders of magnitude. In particular for extensions of the SM with a heavy $Z'$, where direct production at colliders is kinematically suppressed, $e-\tau$ conversion at LHeC provides an exciting new discovery channel for this type of new physics.    
\end{abstract}
\maketitle
\section{Introduction}
Flavor changing neutral current (FCNC) processes in the charged lepton sector are among the most sensitive probes of new physics beyond the current Standard Model (SM) of elementary particles. While they are absent in the SM at tree-level and with vanishing neutrino masses, they do get induced for non-vanishing neutrino masses (via the effective neutrino mass operator) to explain the observed neutrino oscillations at loop level, but only at a level far below foreseen experimental possibilities. 

Extensions of the SM by a heavy neutral gauge boson $(Z^\prime)$ with flavor-violating couplings to the SM fermions provide an interesting scenario of new physics where such FCNC processes are expected to be greatly enhanced. Models of this type can be realized as bottom-up extensions of the SM (see e.g. \cite{NONGUT}) or from GUT theories (see e.g. \cite{GUT}). 

While in many of these models the charged lepton flavor violation (cLFV) with taus is expected to be largest, it is currently much less constrained than cLFV with electrons and muons. Furthermore, regarding ``direct'' collider probes of $Z^\prime$ models, one is limited by the available center-of-mass energy for producing the $(Z^\prime)$ in the $s$-channel, which effectively restricts the searches to the case of $Z^\prime$ masses below a certain mass threshold. 

In this letter, we explore how both of these challenges can be resolved at the LHeC via the $Z'$-mediated $e$-$\tau$ (and $e$-$\mu$) conversion processes $p  e^- \to \tau^- + j$ (and $p  e^- \to \mu^- + j$), where the $Z'$ is exchanged in the $t$-channel.

\section{Effective Lagrangian} 
\label{sec.2}
We consider the low scale effective Lagrangian 
\begin{equation}
\label{eq:1}
{\mathcal{L}_{Z^\prime\bar{f}f}} =\sum_{i,j} Z^\prime \bar{f}_i \gamma^\mu (V^{ij}_L P_L + V^{ij}_R P_R) f_j,
\end{equation}
where $i,j$ run over all fermion degrees of freedom of the SM and $P_{L,R}$ denote the left- and right-chiral projection operators. The parameters $V^{ij}_{L,R}$ parameterize the strength of the $Z^\prime$ coupling to the SM fermions. The Lagrangian in Eq.~(\ref{eq:1}) is generic and includes both flavor-conserving and flavor-violating interactions. 

We note that additional observable effects of this scenario could emerge from gauge kinetic mixing, inducing a mixing of $Z^\prime$ with the $Z$ boson of the SM. This mixing can lead to constraints on the parameters $V^{ij}_{L,R}$ from electroweak precision measurements, and also to cLFV $Z$ decays. However, since we want to focus on the $Z^\prime$ induced cLFV, and since the $Z$-$Z^\prime$ mixing is already constrained by the LEP experiment to be $\le 10^{-3}$ \cite{Abreu:1994ria}, we will ignore these possible effects in the following (and set the mixing to zero).

The current LHC searches for lepton flavor violating heavy neutral gauge boson decays are sensitive to $Z^\prime$ masses up to about $5$ TeV \cite{Sirunyan:2018zhy,Aaboud:2018jff}.
Compared to proton-proton colliders, electron-proton colliders provide an environment where new physics can be probed with comparatively low background rates. 
For our study, we consider the Large Hadron electron Collider (LHeC), which would utilize the $7$-TeV proton beam of the LHC and a $60$-GeV electron beam with up to $80\%$ polarization, to achieve a center-of-mass energy close to $1.3$ TeV with a total of $1 \; ab^{-1}$  
integrated luminosity \cite{AbelleiraFernandez:2012cc,Klein:2009qt,Bruening:2013bga}.

\begin{figure}[h!]
\includegraphics[scale=0.31]{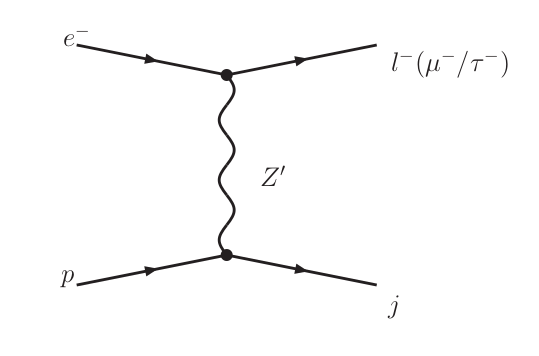}
\caption{Feynman diagram for the $e$-$\tau$ (and $e$-$\mu$) conversion processes $p  e^- \to \tau^- + j$ (and $p  e^- \to \mu^- + j$) mediated by a $Z'$ with flavor-violating couplings to charged leptons at the LHeC.}
\label{F:1}
\end{figure}

As mentioned above, we investigate the LHeC sensitivities to the cLFV $Z^\prime$ couplings via the $e$-$\tau$ (and $e$-$\mu$) conversion processes $p  e^- \to \tau^- + j$ (and $p  e^- \to \mu^- + j$) mediated by a $Z'$ with lepton flavor-violating couplings in the $t$-channel. The matrix elements of these processes, with the Feynman diagram shown in Fig.~\ref{F:1}, are sensitive to the cLFV parameters $V_{L,R}^{e \mu}$ and $V_{L,R}^{e \tau}$ from the $Z^\prime$ coupling to the leptons, as well as to the couplings of the $Z^\prime$ to the constituent quarks of the proton and the one that leads to the final state jet. 

To compare the LHeC sensitivity with the current experimental limits from searches for flavor-conserving and flavor-violating processes, we will set the couplings $V^{ij}_{L,R}$ to be equal for all channels, i.e.\ $V^{ij}_L=V^{ij}_R =: V$ for all $i,j$. We like to emphasize that for a specific model, the individual limits as well as the LHeC sensitivities can be reconstructed by scaling the result with the combination of the $V^{ij}_{L,R}$ the respective process depends on.

\begin{figure}[h!]
\includegraphics[scale=0.22]{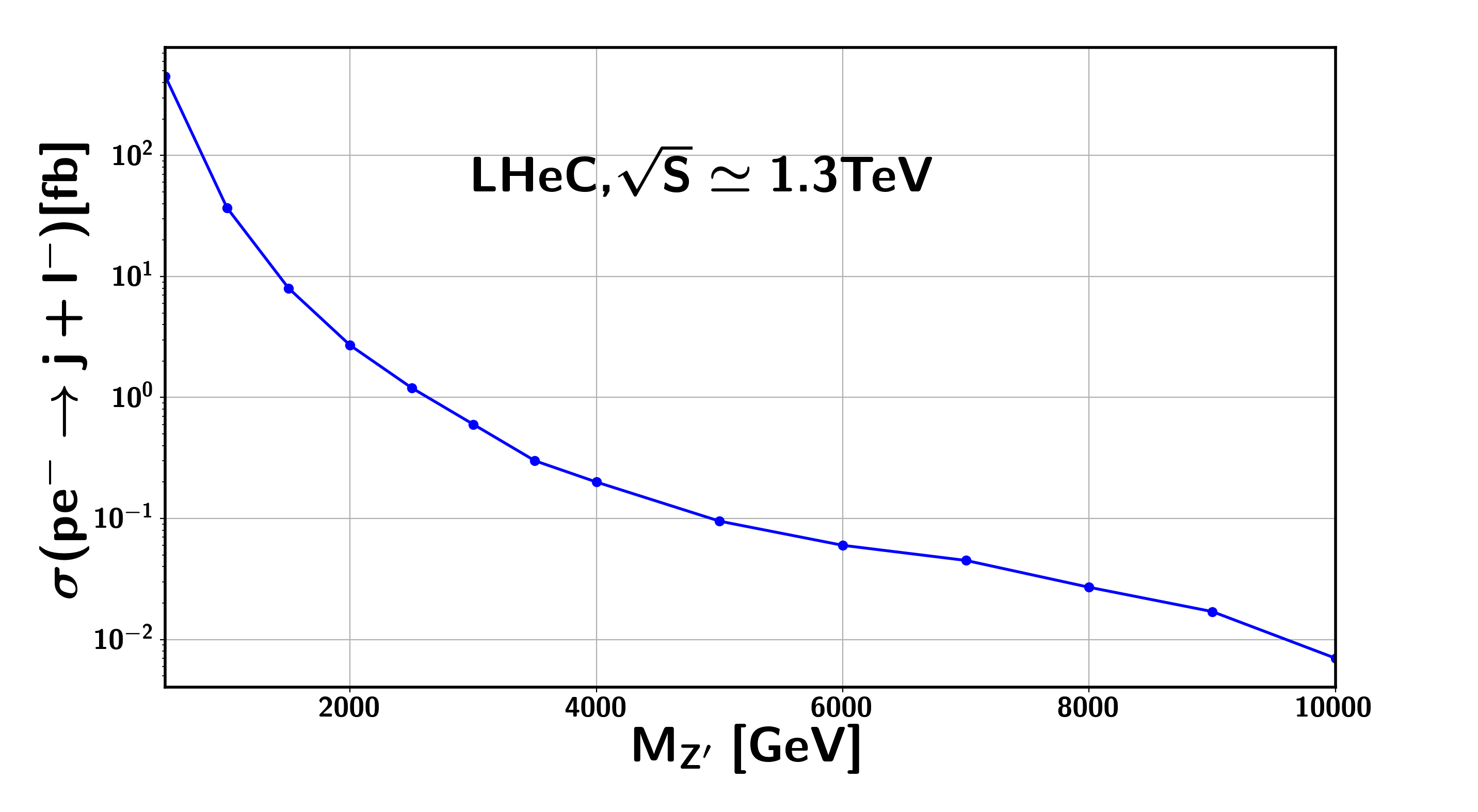}
\caption{Total cross section for the $Z'$-mediated processes $p  e^- \to l_\alpha^- + j$ with $\alpha = \mu,\tau$ at the LHeC for the example value $V = 0.1$.}
\label{F:2}
\end{figure}

The total cross section for the $Z'$-mediated processes $p  e^- \to l_\alpha^- + j$ with $\alpha \not= e$ scales as $|V|^4$. It is shown in Fig.~\ref{F:2} for the example value $V = 0.1$ as a function of the $Z'$ mass $(M_{Z'})$.

 In the following, we focus on the LHeC sensitivity for the $Z'$-mediated $e$-$\tau$ conversion processes $p  e^- \to \tau^- + j$, and we will later comment on the 
$e$-$\mu$ conversion process $p  e^- \to \mu^- + j$. 

\section{NON-COLLIDER EXPERIMENT CONSTRAINTS}
\label{sec.3}

In this section, we consider constraints on the $Z^\prime$ coupling strength parameter $|V|^2$ from non-collider experiments with taus, where the most relevant current constraints on the parameters $V_{L,R}^{e \tau}$ come from two- and three-body tau decays. Note that, as explained above, we will below set the couplings $V^{ij}_{L,R}$ to be equal for all involved channels, i.e.\ $V^{ij}_L=V^{ij}_R =: V$ for all $i,j$, to allow for a simple comparison of the strength of the various experimental sensitivities.

\subsection{Two body decays of tau leptons}
\label{sec.3}

The decay rate of $\tau \rightarrow e \gamma$ is given by \cite{Lavoura:2003xp,Chiang:2011cv,Lindner:2016bgg,Raby:2017igl},
\begin{equation}
	\Gamma(\tau\rightarrow e\gamma)=\frac{\alpha_{em}}{1024\pi^{4}}\frac{m_{\tau }^{5}}{M_{Z^{\prime }}^{4}}(\left\vert \widetilde{\sigma }_{L}\right\vert ^{2}+\left\vert \widetilde{\sigma }_{R}\right\vert^{2}),
\label{eqn:mu_e_gamma_decay_rate_prediction}
\end{equation}
with $\widetilde{\sigma }_{L}$ and $\widetilde{\sigma }_{R}$ defined as
\begin{align}
	\begin{split}
	\widetilde{\sigma }_{L}& =V^2 \sum_{a=e,\mu,\tau}\left[ F(x_{a})+\frac{m_{a}}{m_{\tau }} G(x_{a})\right] , \\
	\widetilde{\sigma }_{R}& = V^2 \sum_{a=e,\mu,\tau}\left[  F(x_{a})+\frac{m_{a}}{m_{\tau }} G(x_{a})\right].
	\label{eqn:contributions_to_muegamma_(sigmas)}
	\end{split}
\end{align}
$m_{a}$ (with $a\in\{e,\mu,\tau\}$) are the charged lepton masses, $x_{a}=m_{a}^{2}/M_{Z^{\prime }}^{2}$ and $F(x)$ and $G(x)$ are the respective loop functions,
\begin{eqnarray}
F(x)&=& \frac{5x^{4}-14x^{3}+39x^{2}-38x-18x^{2}\ln x+8}{ 12(1-x)^{4}},\\ 
G(x)&=&\frac{x^{3}+3x-6x\ln x-4}{2(1-x)^{3}}. 
\label{eqn:loop_functions2}
\end{eqnarray}
The experimental limit on the branching ratio ${\rm BR}(\tau \to e \gamma) = \Gamma(\tau\rightarrow e\gamma) / \Gamma_{\tau}$, where $\Gamma_{\tau}$ is the total tau decay width, is given by $3.3 \times 10^{-8}$ at $90\% $ confidence level  \cite{Aubert:2009ag}.

\subsection{Three body decays of tau leptons}
\label{sec.3}

The branching ratio of $\tau \to l_i l_j \bar{l}_k$ takes the form  \cite{Langacker:2000}
\begin{eqnarray}
{\rm BR}(\tau \to l_i \, l_j \,  \bar{l}_k) &=& \frac{m_{\tau}^5}{1536  \, \pi^3  \, \Gamma_{\tau} }\left(\left| {C_{L}^{3l}}_{ijk} + {C_{L}^{3l}}_{jik} \right|^2 \right. \nonumber\\
&& \left. +\left| {C_{R}^{3l}}_{ijk} \right|^2 + \left| {C_{R}^{3l}}_{jik} \right|^2 \right),
\end{eqnarray}
with the coefficients given by
\begin{eqnarray}
 C^{3l}_{L}  &=&  \left \{ \frac{1}{\Lambda^2_{Z'}}  - \frac{ \cos 2 \theta_W }{2} \frac{1}{\Lambda^2_Z} \right \}, \\
  C^{3l}_{R} &=&  \left \{  \frac{1}{\Lambda^2_{Z'}}  +  \sin^2 \theta_W  \frac{1}{\Lambda^2_{Z}}  \right \},
\end{eqnarray}
where
\begin{equation}
\label{eq;C1}
\frac{1}{\Lambda_{Z'}^2} =  \left( \frac{V^2\cos^2 \theta}{M^2_{Z'}} +\frac{V^2 \sin ^2 \theta}{M^2_{Z}}  \right ),
\end{equation}
\begin{equation}
 \frac{1}{ \Lambda^2_{Z}}= V g_Z \sin \theta \cos \theta  \left ( \frac{1}{M^2_{Z}}-\frac{1}{M^2_{Z'}} \right ), \; \text{and}
\end{equation}
\begin{equation}
\label{eq;ZZpmixing}
\tan 2 \theta \simeq 4 \, \frac{V}{g_Z} \frac{M^2_{Z}}{ M^2_{Z'}}.
\end{equation}

The current experimental bound on the branching ratio $\tau \rightarrow 3e$ is $2.7\times 10^{-8}$ at $90\% $ confidence level  \cite{Hayasaka:2010np}.

\section{Bounds from direct searches at the LHC}
\label{sec.3}

In addition to the (indirect) limits from non-collider experiments, we consider constraints from direct searches at the LHC. 
In particular, in order to compare with the sensitivity of $e$-$\tau$ ($e$-$\mu$) conversion at the LHeC, we consider the limits from LHC searches for $Z^\prime$ decays into $e  \tau$ ($e  \mu$) pairs \cite{Aaboud:2018jff}. The considered searches have total integrated luminosity of $36.1\; \text{fb}^{-1}$ and center-of-mass energy of 13 TeV. With no excess over the SM predictions observed, limits have been placed on the $Z^\prime$ mass and its coupling strength at the $95\%$ confidence level. 

Furthermore, we also consider the LHC search for $Z^\prime$ decays into same-flavor dielectron and dimuon states \cite{Aad:2019fac}, which currently give the strongest collider constraints on $Z^\prime$ parameters. The search has total integrated luminosity of $139\; \text{fb}^{-1}$ and center-of-mass energy of 13 TeV in the mass range between 250 GeV to 6 TeV. No deviation from the Standard Model predictions has been observed, leading to an upper limit on the fiducial cross-section times branching ratio at the $95\%$ confidence level. The limit can be converted into a constraint on the mass of the $Z^\prime$ and its coupling strength (which we parameterize by $|V|^2$). 

For comparison, we will include the limits on $|V|^2$ from these searches and from the most non-collider experiments most sensitive to $e$-$\tau$ ($e$-$\mu$) flavour transitions in Fig.~\ref{FF}, together with the LHeC sensitivities to be discussed in the next section.


\section{LHeC sensitivity }
\label{sec.4}
In this section, we discuss the sensitivity of the LHeC to the cLFV $e$-$\tau$ conversion process
\begin{equation}
pe^- \to j + \tau^- \: ,
\label{eq:3.2}
\end{equation}
mediated by a $Z'$ with lepton flavor-violating couplings in the $t$-channel. As mentioned earlier, the t-channel process has a comparatively weak dependence on the $Z^\prime$ mass, and  its differential cross section relies on the kinematics of the boosted tau lepton. The process is absent in the SM and provides a powerful search tool for new physics. 

The dominant source of background stems from SM gauge boson decays or radiated soft taus. For tau lepton reconstruction, we used an identification efficiency rate $75\%$ for tau leptons with $P_T\ge 40$ GeV and miss-identification rate about $1\%$ \cite{Bagliesi:2007qx,Bagliesi:2006ck}.  The most relevant backgrounds and their  total cross sections are shown in table \ref{tab:1}.
\begin{table}[h!]
\begin{center}
\begin{tabular}{|l|c|}
\hline
Backgrounds & $\sigma_{(LHeC)} [Pb]$   \\ \hline\hline
$p e^-\to Z \ \nu_l\  j , \quad \mbox{where}  \;Z\to \tau^-\tau^+$  & 0.0316   \\ \hline
$p e^-\to  W^\pm \ e^-\  j , \quad \mbox{where}  \;W^\pm\to \tau^\pm\ {\nu}_\tau$  &0.2657  \\ \hline
$p e^-\to Z Z \ \nu_l\  j , \quad \mbox{where}  \;Z\to \tau^-\tau^+$  & 1.1$\times 10^{-5}$   \\ \hline
$p e^-\to Z W^\pm \ \nu_l\  j $,& \\  $ \quad\mbox{where}  \;Z\to \tau^-\tau^+ ,\;W^\pm\to \tau^\pm\ {\nu}_\tau$  & 2.64$\times 10^{-5}$   \\ \hline
\end{tabular}
\end{center}
\caption{Dominant background processes considered in our analysis and their total cross sections. The samples have been produced with the following cuts: $P_T(j)\ge 5$ GeV, $P_T(l)\ge 2$ GeV and $|\eta(l/j)|\le 4.5$. }
\label{tab:1}
\end{table}

It is worth mentioning that other backgrounds like $pe^-\to h\ \nu_l\ j$ with the SM Higgs $h$ decaying to a tau pair is suppressed by the small electron Yukawa coupling, while the process of single top production $pe^-\to\nu_l\  t$ is suppressed by the small involved CKM mixing matrix element. 

For the analysis and to distinguish between the signal events and all relevant backgrounds, we have constructed $31$ kinematic variables (at the reconstruction level after the detector simulation) which are used as input to the Tool for Multi-Variate Analysis (TMVA). The Machine Learning algorithm Boosted Decision Trees (BDT) is used to separate the signal events from the background events as in Ref.~\cite{Antusch:2019eiz}. 

\begin{figure}[h!]
\includegraphics[scale=0.17]{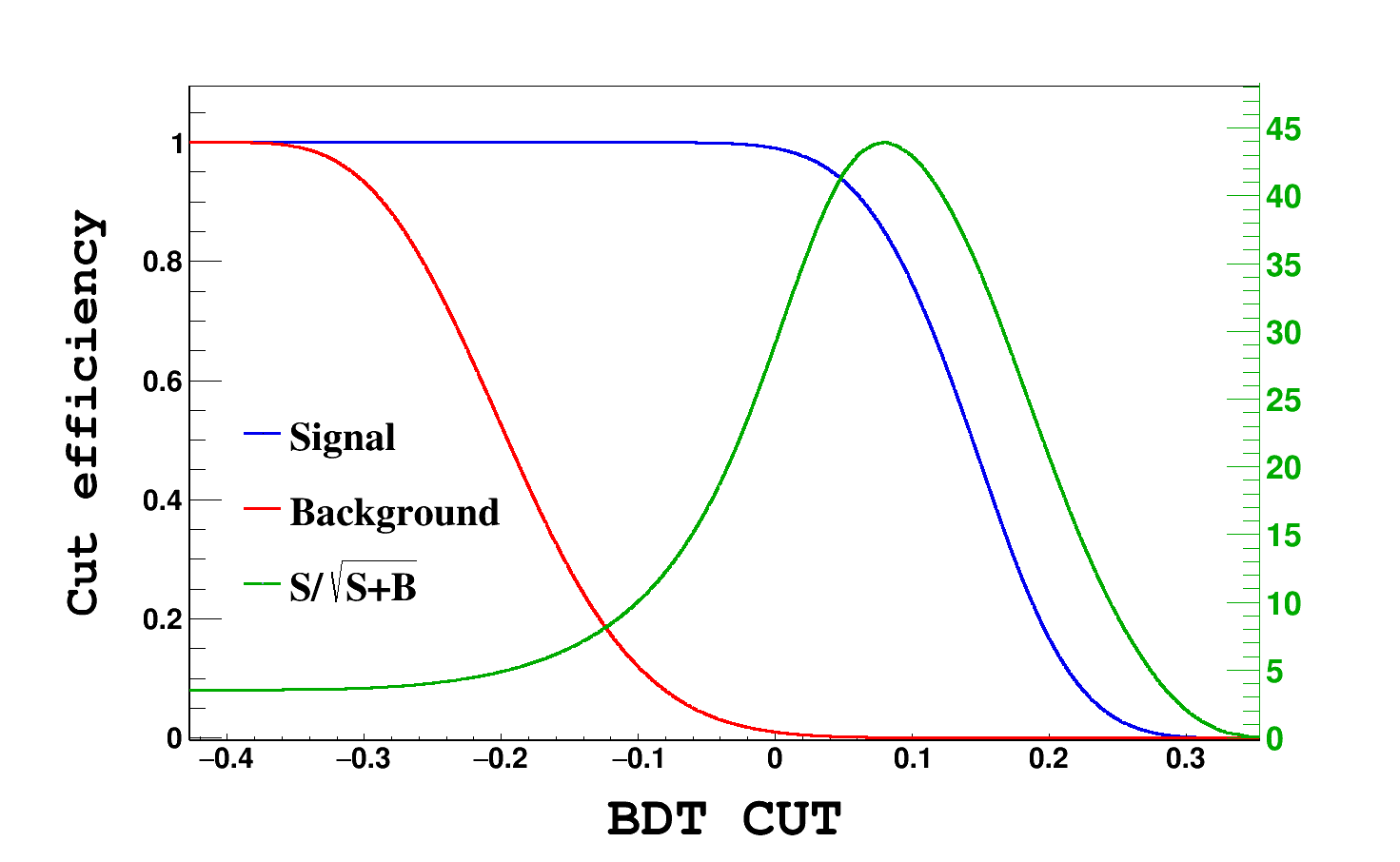}
\caption{Cut efficiency at the LHeC with BDT cut $\ge 0.081$. One can get $S/\sqrt{S+B} = 42.2\sigma$ with number of signal events $=1994$, and background events $=230$. The cut efficiency for the signal is $0.85$, and for the background is $5\times 10^{-4}$. The benchmark point is chosen with $M_{Z^\prime} = 2$ TeV and $V=0.1$. }
\label{F:3}
\end{figure}

The BDT rank shows that the most important variable for discriminating the signal events from the background events  is the tau transverse momentum. 
However, the other variables like the invariant mass of the tau lepton pair, the transverse mass of the tau lepton, the missing energy, the transverse momentum of electrons and positrons, $\Delta R$ between tau lepton and the beam jet, and $\Delta R$ between tau lepton and electron are all of similar importance for the separation of signal and background. This indicates that our signal process has a characteristic behavior that can be easily distinguished from the relevant backgrounds. For illustrative purpose, we show the optimization of the signal significance as a function of signal and background cut efficiency for a selected benchmark point in Fig.~\ref{F:3}.
\\

\section{Results}
\label{sec.5}
\begin{figure*}[th!]
\centering
\includegraphics[scale=0.4]{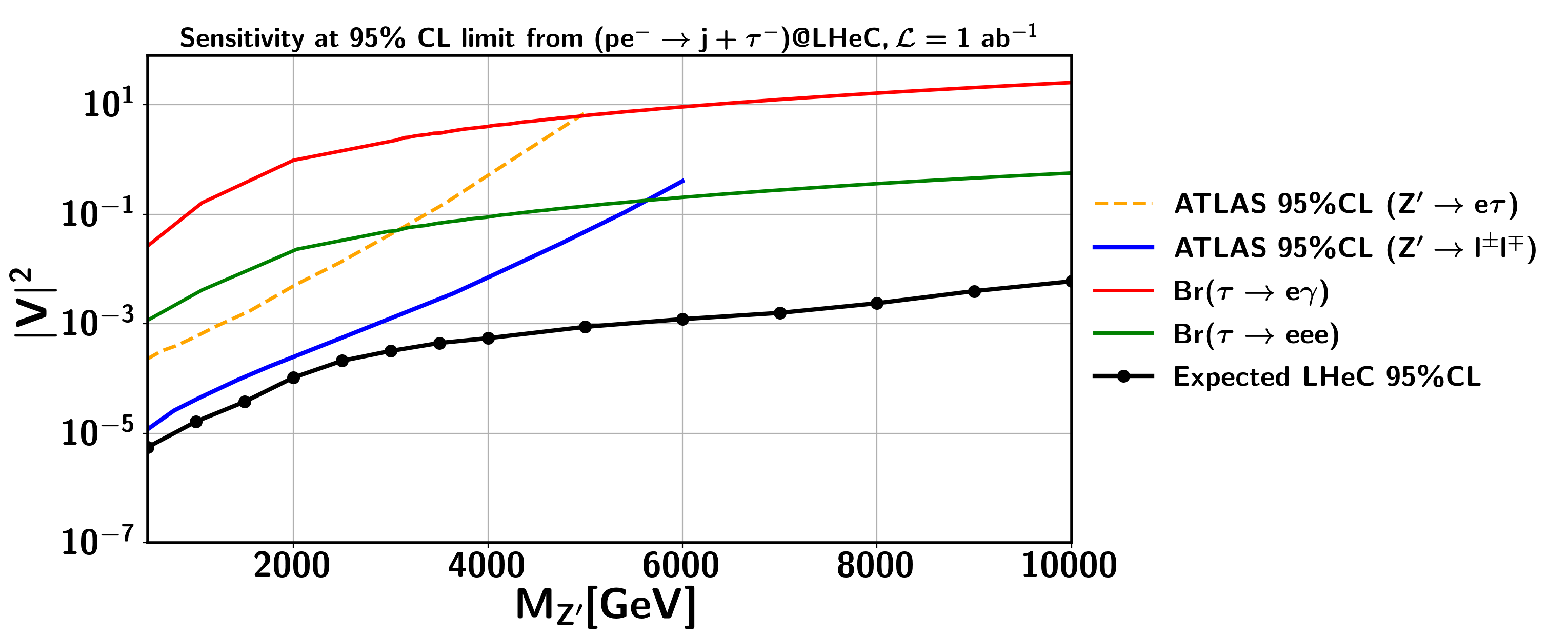}\\
\includegraphics[scale=0.4]{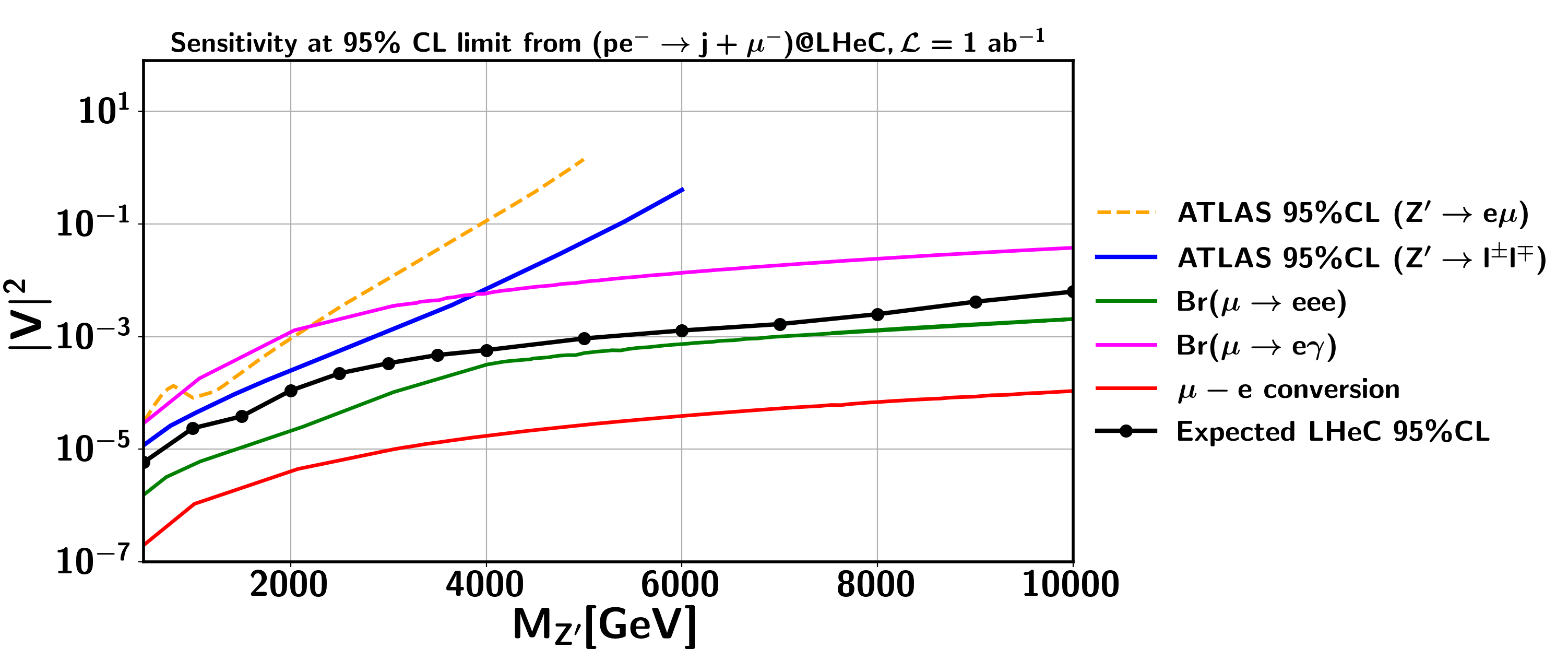}
\caption{Expected limits on the parameter $|V|^2$ when testing the signal hypothesis, setting $V^{ij}_L=V^{ij}_R=:V$ as described in the main text to compare with the existing limits from experimental constraints on the relevant flavor conserving and flavor violating processes. The black line is the main result of our analysis, i.e.\ the forecast of the LHeC sensitivity at $\sqrt{S}=1.3$ TeV and integrated luminosity of $1\ ab^{-1}$ with including $2\%$ systematic uncertainty. {\it{Upper Panel}}: Expected LHeC limit on the parameter $|V|^2$ for the process $pe^-\to\tau j$ with all the best sensitivity limits from the current  collider and non-collider searches. {\it{Lower Panel}}: Expected LHeC limit on the parameter $|V|^2$ for the process $pe^-\to\mu j$ with all the best sensitivity limits from the current collider and non-collider searches.}
\label{FF}
\end{figure*}

Given the number of signal events and the number of background events after the BDT optimized cuts, the LHeC limit at $95\%$ confidence level is obtained using the formula \cite{Antusch:2018bgr}:
\begin{equation}
\resizebox{0.475\textwidth}{!}{$\sigma_{sys} = \left[2\left((N_s+N_b) \ln\frac{(N_s+N_b)(N_b+\sigma_b^2)}{N^2_b+(N_s+N_b)\sigma^2_b} - \frac{N^2_b}{\sigma^2_b}\ln (1+\frac{\sigma^2_b N_s}{N_b(N_b+\sigma^2_b)} \right)\right]^{1/2}$},
\end{equation}
with $N_s$ and $N_b$ being the number of signal and background events, and $\sigma_b$ is the systematic uncertainty, taken to be $2\%$ for background events only. In Fig.~\ref{FF} (upper plot), we show the LHeC sensitivity on $|V|^2$ via the $e$-$\tau$ conversion process $p  e^- \to \tau^- + j$ (black line). 
For comparison, we also show the most recent limits from the most sensitive collider and non-collider experiments assuming, as stated above, equal $Z^\prime$ couplings for all flavor violating and conserving decay channels to fermions.  

In this context, the LHC  searches for lepton flavor violating or lepton flavor conserving $Z^\prime$ decays are very sensitive  in the $Z^\prime$ mass range from $500$ GeV to $3$ TeV, while for larger masses the sensitivity drops strongly. The reason for this drop is that the $Z^\prime$ production at the LHC is mainly via the  s-channel, with the $Z^\prime$ produced on the mass shell. This means the kinematic restrictions strongly limit the mass reach. 

The non-collider limits from the two and three body decays of tau lepton are not as strong in the mass range from $500$ GeV to $3$ TeV, while for larger masses  they become more sensitive than the LHC searches. The LHeC sensitivity can be best in the whole mass region we considered (above $500$ GeV). 

For completeness, we also discuss the LHeC sensitivity via the $Z^\prime$-mediated $e$-$\mu$ conversion process $p  e^- \to \mu^- + j$.  
The results are shown in Fig.~\ref{FF} (lower plot) along with the current limits from the most relevant collider and non-collider experiments (where we have also included the very strong constraints from $\mu-e$ conversion in nuclei).  
The sensitivity we obtain is similar to the one for the tau process, since we have assumed equal $Z^\prime$ couplings  ($=V$) to all fermion pairs. Also the dominant  backgrounds include the ones in table \ref{tab:1}, replacing the tau with muon. Moreover, we include additional backgrounds for soft muons that come from the leptonic tau decays.  We can see that the current LHC and $\mu\to e\gamma$ limits \cite{Lavoura:2003xp,Chiang:2011cv,Lindner:2016bgg,Raby:2017igl,CarcamoHernandez:2019ydc,TheMEG:2016wtm}  are comparatively weak (compared to the LHeC sensitivity estimate), while the bound from $\mu \to eee$ \cite{Hisano:2016afc,Bellgardt:1987du,Blondel:2013ia} and, in particular, $\mu-e$ conversion in nuclei \cite{Hisano:2016afc,Kitano:2002mt,Bertl:2006up,Galli:2019xop} give the best sensitivities for cLFV $Z^\prime$ couplings with final state muon.
This means that, as expected, the LHeC sensitivity for the  $e$-$\mu$ conversion process $p  e^- \to \mu^- + j$ cannot exceed the very strong sensitivities of the present searches for cLFV involving electrons and muons. 

On the other hand, the LHeC sensitivity to the $e$-$\tau$ conversion process
$pe^- \to j + \tau^-$ can exceed the current sensitivities by more than two orders of magnitude (for heavy $Z^\prime$ above about $3$ TeV), allowing for interesting discovery possibilities.

\section{Conclusions}
\label{sec.6}
In this letter, we have studied the sensitivity of the LHeC to $e$-$\tau$ (and $e$-$\mu$) conversion processes $p  e^- \to \tau^- + j$ (and $p  e^- \to \mu^- + j$) mediated by a $Z^\prime$ with lepton-flavor violating couplings in the $t$-channel. The results are presented in Fig.~\ref{FF}, where we have parameterized the $Z^\prime$ couplings to fermions by the general Lagrangian of Eq.~(\ref{eq:1}) and used equal $Z^\prime$ couplings (i.e.\ $V^{ij}_L=V^{ij}_R =: V$) for all channels to give an explicit example and to compare with existing bounds. Using these results, the LHeC sensitivities as well as the current limits can be obtained for a specific model (with model-dependent $V^{ij}_L$, $V^{ij}_R$) by scaling the results with the combination of the $V^{ij}_{L,R}$ the respective process depends on.  

Compared to current tests at the LHC, where cLFV decays of the $Z^\prime$ (produced in the s-channel) are searched for, the LHeC has sensitivity to much higher $Z^\prime$ masses, up to {\cal O}(10) TeV. For cLFV with taus, we find that the LHeC sensitivity from the process $p  e^- \to \tau^- + j$ can exceed the current limits from collider and non-collider experiments in the considered $Z^\prime$ mass range (above $500$ GeV) by more than two orders of magnitude. In particular for extensions of the SM with a heavy $Z^\prime$ (above about $3$ TeV), where direct production at colliders is kinematically suppressed, lepton flavor conversion with taus at the LHeC offers exciting discovery prospects for this type of new physics beyond the SM.    

\section*{Acknowledgments}
This work has been supported by the Swiss National Science Foundation. A.R. acknowledges the hospitality of the Department of Physics, University of Basel  where the visit was supported through the SU-FPDC Grant Program.

\end{document}